\documentclass[twocolumn,showpacs,amsmath,amssymb,floatfix,aps]{revtex4}
\usepackage{graphicx}
\usepackage{dcolumn}
\usepackage{bm}
\begin{document}
\title{Potential Energy Landscape Equation of State}
\author{Emilia La Nave$^{1}$, Stefano Mossa$^{2,1}$, 
and Francesco Sciortino$^{1}$}
\affiliation{$^{1}$ 
Dipartimento di Fisica, INFM UdR  and INFM Center for
Statistical Mechanics and Complexity, 
Universit\`{a} di Roma ``La Sapienza'', Piazzale Aldo
Moro 2, I-00185, Roma, Italy\\
$^{2}$ Center for Polymer Studies and Department of Physics\\
Boston University, Boston, MA 02215 USA}
\date{\today}
\begin{abstract}
Depth, number, and shape of the basins of the 
potential energy landscape 
are the key ingredients of the inherent structure thermodynamic 
formalism introduced by Stillinger and Weber~[F. H. Stillinger 
and T. A. Weber, Phys. Rev. A {\bf 25}, 978 (1982)]. 
Within  this formalism, an equation of state based only on 
the volume dependence of these landscape properties is derived.
Vibrational 
and configurational contributions to
pressure are sorted out in a transparent way.
Predictions are successfully compared with data 
from extensive molecular dynamics simulations
of a simple model for the  fragile liquid orthoterphenyl.
\end{abstract}
\maketitle
Recent years have seen a resurgence in studies devoted to
modeling the thermodynamics of supercooled 
liquids~\cite{thermo4,thermo3,angell,thermo2,wales}.
Such studies aim  to elucidate the physics of the liquid-glass transition,
to develop a thermodynamic description of out of
equilibrium systems and to provide keys for 
a deeper understanding of the dynamics of supercooled 
states~\cite{debenedetti01}. 
Numerical studies, boosted by increased computational 
resources which now allow simulations to track the
slowing down of the dynamics over more than 6 decades in time,
are  providing quantitative estimates for the 
free energy of simple model systems \cite{lennard_jones_pes,heuer97,
scala2000,sastry2001}.
The availability of such data provides stringent tests of 
the theoretical predictions~\cite{scala2000,sastry2001,ivannature}
and helps in the understanding of basic mechanisms associated with
the behavior of thermodynamic and dynamic quantities close to the 
glass transition.

Among the thermodynamic formalisms amenable to
numerical investigation, a central role is played by the
Inherent Structure (IS) formalism introduced 
by Stillinger and Weber~\cite{sw}. Properties of the
potential energy landscape (PEL), such as depth, number  
and shape of the basins of the potential energy surface
are calculated and used in the evaluation of the
liquid free energy~\cite{scala2000,sastry2001,ivannature,mossa02}
In the IS formalism, the system free energy 
is expressed as a sum of an entropic contribution, 
accounting for the number of the available basins, 
and a vibrational contribution,
expressing the free energy of the system when constrained in one 
of the basins~\cite{sw}.

Important progress has been made after the discovery 
that --- for models of fragile liquids ---
the number $\Omega(e_{IS})$ of distinct basins of depth $e_{IS}$ 
in a system of $N$ atoms or molecules is well
described by a Gaussian distribution~\cite{sastry2001,heuer00} 
\begin{equation}
\Omega(e_{IS})=e^{\alpha N} 
\frac{e^{-(e_{IS}-E_o)^2/2\sigma^2}}{(2 \pi \sigma^2)^{1/2}}.
\label{eq:gdos}
\end{equation}
Here the amplitude $e^{\alpha N}$ accounts for the total number of
basins.  Numerical studies of models for fragile liquids 
have also shown that the basin free energy can be written 
as the depth $e_{IS}$ plus a vibrational contribution which, 
in the harmonic approximation, has the well known form
\begin{equation}
F_{vib}(e_{IS},T)= 
k_B T \sum_{i=1}^{M}   
\ln(\beta\hbar\omega_i(e_{IS})),  
\label{eq:fbasin}
\end{equation}
where $\omega_i(e_{IS})$ is the $i$-th normal mode frequency  ($i=1...M$)
and $\beta=1/k_BT$.  The $M$ normal mode frequencies 
define the shape of the basin. If relevant, 
anharmonic corrections can also be accounted 
for~\cite{ivannature,mossa02}. The quantity 
$\sum_{i=1}^{M} \ln( \omega_i(e_{IS})/\omega_o)$
(where $\omega_o$ is the frequency unit) 
is found to depend linearly on the basin 
depth~\cite{sastry2001}, i.e.
can be written, in terms of two parameters $a$ and $b$, as
\begin{equation}
\sum_{i=1}^{M} \ln( \omega_i(e_{IS})/\omega_o) = a +  b \, e_{IS}.
\label{eq:logw}
\end{equation}
Hence, the vibrational free energy can be written as
\begin{equation}
F_{vib}(e_{IS},T) = F_{vib}(E_o,T) + k_B T \, b (e_{IS} - E_o).
\label{eq:fbasin2}
\end{equation}
Within the two assumptions of Eq.~(\ref{eq:gdos}) ---
Gaussian distribution of basin depths --- and  Eq.~(\ref{eq:fbasin2}) ---
linear dependence of the basin free energy on  $e_{IS}$ --- 
an exact evaluation of the partition function can be carried out. 
The corresponding Helmholtz free energy is given by~\cite{sastry2001} 
\begin{eqnarray}
\label{eq:freeenergy}
F (T)&=&-T S_{conf}(T)+\langle e_{IS}(T)\rangle \\
&+ &F_{vib}(E_o,T) +k_B T \, b(\langle e_{IS}(T)\rangle -E_o),\nonumber
\end{eqnarray}
where 
\begin{equation}
\langle e_{IS}(T)\rangle =(E_o-b\sigma^2)-\beta \sigma^2= 
e_\infty-\beta \sigma^2,
\label{eq:eis}
\end{equation}
and
\begin{eqnarray}
S_{conf}(T)/k_B&=& \alpha N 
-(\langle e_{IS}(T)\rangle -E_o)^2/2 
\sigma^2\nonumber\\ 
&=&S_\infty/k_B- b\,\beta\,\sigma^2-\beta^2\sigma^2/2.
\label{eq:sconf2}
\end{eqnarray}
In the above equations, $e_\infty$ and $S_\infty$ 
are defined as the value of $\langle e_{IS}\rangle$ 
and $S_{conf}$ at infinite $T$.
Eqs.~(\ref{eq:freeenergy})-(\ref{eq:sconf2}) show that, along constant
volume $V$ paths, the behavior of the thermodynamic quantities is
controlled by the values of the PEL properties,
as given by $\alpha$, $\sigma$, $E_0$ (from Eq.~(\ref{eq:gdos})) 
and by $a$ and $b$ (from Eq.~(\ref{eq:logw})). 
\begin{figure}[t] 
\centering
\includegraphics[width=.38\textwidth]{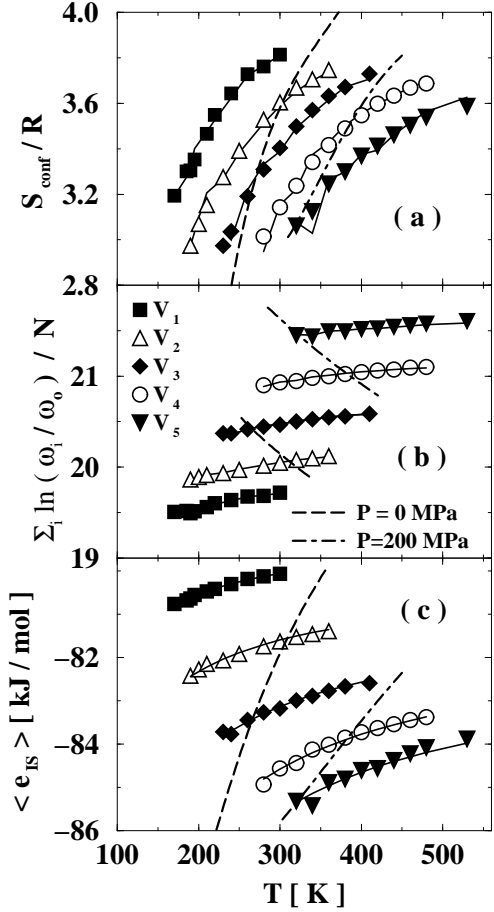}
\vspace{-0.5cm}
\caption{
T dependence of (a) $S_{conf}$, (b)  $ \sum_i \ln(\omega_i/\omega_o)/N$, 
and (c) $\langle e_{IS}\rangle$ (per molecule) 
at the five studied volumes $V_k$ (symbols). 
Data are from Ref.~\protect\cite{mossa02}. 
The full curves are simultaneous fits of the three sets of data according
to Eqs.~(\protect\ref{eq:sconf2}), (\protect\ref{eq:logw}), 
and (\protect\ref{eq:eis}).
The dashed curves are the constant $P$ paths (at $P=0$ and $200$ MPa)
calculated according to the IS-EOS as discussed in the text.
The frequency unit is $\omega_0\equiv 1 cm^{-1}$}
\label{fig:fit}
\end{figure}

In this Letter we study the volume dependence of 
Eq.~(\ref{eq:freeenergy})
to provide an expression for the equation of state (EOS) based 
completely
on landscape properties (PEL-EOS) \cite{debenedettieos}.
This study provides a significant insight 
into the understanding of the pressure
$P$ and opens the way for detailed comparisons 
between experimental measurements
(usually performed at constant $P$) and theoretical
approaches based on the IS formalism.  
It may also help in developing an IS-based 
thermodynamic description of out-of-equilibrium 
(glass) states and a theoretical definition of the concepts of 
fictive $P$ and $T$\cite{aging}.
\begin{figure}[t] 
\centering
\includegraphics[width=.5\textwidth]{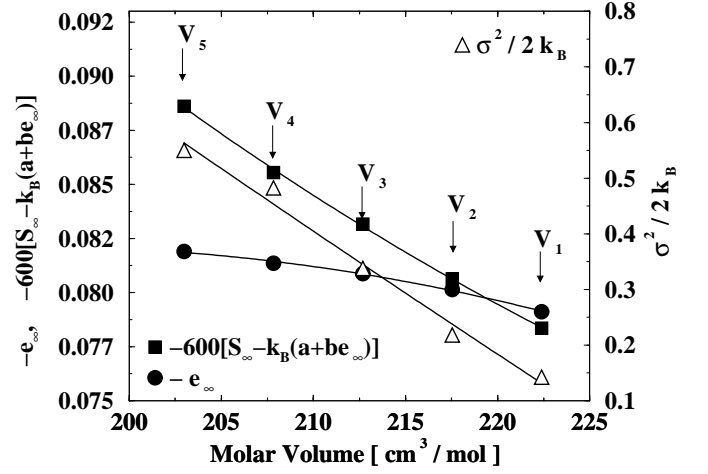}
\vspace{-0.5cm}
\caption{
$V-$dependence of the three combinations of PEL parameters 
contributing to the linear (squares), constant (circles)
and $T^{-1}$ (triangles) components of $P$, 
according to Eq.~(\protect\ref{eq:componenti}).
The solid lines are the polynomial fit used to evaluate 
$P_{const}$, $P_{_T}$ and $P_{1/T}$.
Tables of the coefficients of the polynomial fit will be  reported in
Ref.~\protect\cite{lanave02}. Here $e_{\infty}$ is expressed in
$10^6 J /mol$, $S_\infty-k_B(a+be_\infty)$ in $10^6 J/mol/K$, and 
$\sigma^2/2 k_B $ in $10^6 J K /mol$.
\label{fig:fitp}
}
\end{figure}

In thermodynamics, $P$ is defined as the (negative)
$V$ derivative of the Helmholtz free energy. Hence 
$P$ if fully determined by the  $V$ dependence 
of the landscape properties $\alpha$, $\sigma$, $E_0$, $a$ and $b$.  
Eq.~(\ref{eq:freeenergy}) shows that $P$ can be split into three main 
contributions: a configurational one, $P_{conf}$ --- 
related to the change in the number of available basins with $V$;  
an $e_{IS}$ one, $P_{e_{IS}}$ --- related to the
change in basin depth with $V$; and
a vibrational one, $P_{vib}$ --- related to the change in 
the shape of the explored basin with $V$. The $T$ dependence of
each contribution can then be studied independently.
The explicit expressions for these contributions are:
\begin{eqnarray}
P_{conf}(T,V)= 
T \frac {\partial} { \partial V} S_\infty -
\frac {\partial} { \partial V} (b\sigma^2) -
\frac{1}{T}   \frac {\partial} { \partial V} (\sigma^2/2k_B)
\label{eq:pconf}
\end{eqnarray}
\begin{equation}
P_{e_{IS}}(T,V)= 
-\frac{\partial}{\partial V } e_\infty
+ 
\frac{1}{T} 
\frac{\partial}{\partial V } (\sigma^2/k_B)
\end{equation}
\begin{equation}
P_{vib}(T,V)= 
- k_B T \frac{\partial}{\partial V } (a+be_\infty) +
\frac{\partial}{\partial V } (b\sigma^2).
\label{eq:pvib}
\end{equation}
By grouping together all the contributions with the same
$T$ dependence, $P$ can be expressed in term of $V$ derivatives 
of only three combinations of the five PEL parameters\cite{anomali} 
\begin{equation}
P(T,V)= P_{const}+T P_{_{_{T}}}+T^{-1} P_{_{_{1/T}}},
\label{eq:componenti}
\end{equation}
where $P_{const}=- \partial e_{\infty} /\partial V$, 
$P_{_{_{T}}}=\partial S_\infty /\partial V - k_B 
 \frac {\partial} { \partial V} (a+be_\infty)$ 
and $P_{_{_{1/T}}} = \frac {\partial} { \partial V} (\sigma^2/2k_B)$.

The present formalism also provides an expression 
for the so-called inherent structure equation of state (IS-EOS),
$P_{IS}(T_{IS},V)$~\cite{debenedettieos,laviolette,roberts,sastry00}, 
i.e. the relation between  the  pressure and  volume of the typical IS
and the temperature $T_{IS}$ of the equilibrium 
ensemble from which configurations were extracted.
Indeed, the constant $V$ steepest descent minimization procedure, which
realizes the search for the closest local minimum 
starting from an equilibrium configuration, 
suppresses all the vibrational components (hence $P_{vib}=0$)
but keeps $P_{conf}$ and $P_{e_{IS}}$ frozen at their 
``equilibrium'' initial value. As a result  $P_{IS}$,
a purely mechanical quantity, can be expressed 
as
\begin{eqnarray}
P_{IS}(T_{IS},V)=P_{conf}+P_{e_{is}} = \nonumber \\
-\frac{\partial}{\partial V} E_o + 
T_{IS}   \frac{\partial}{\partial V}  S_\infty + 
\frac{1}{k_B T_{IS}}  
\frac{\partial}{\partial V}  (\sigma^2/2).
\label{eq:pis}
\end{eqnarray}
\begin{figure}[t] 
\centering
\includegraphics[width=.45\textwidth]{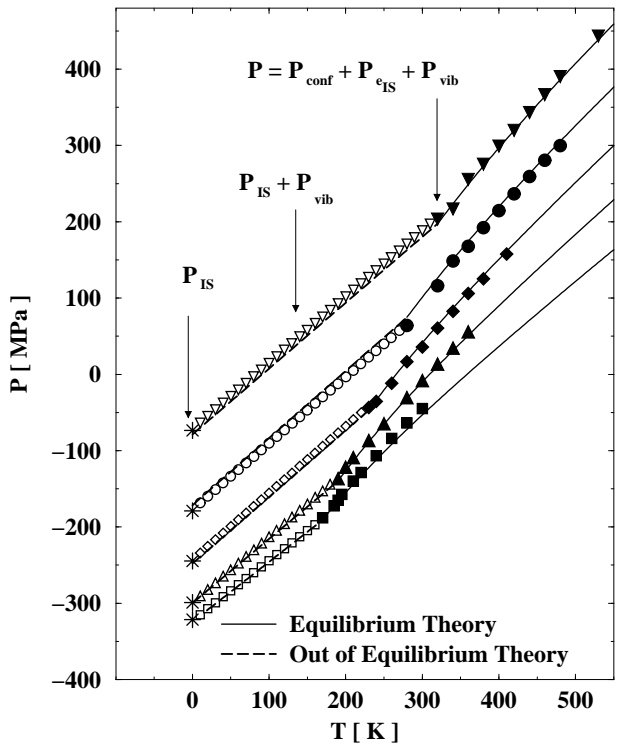}
\vspace{-0.5cm}
\caption{
Comparison between $P$ evaluated according to
the $V$ derivative of PEL properties (lines) and $P$ 
evaluated in the MD simulation using the virial expression 
(symbols). 
Solid symbols are equilibrium values. For each $V_k$,  
the symbol $*$ indicates $P_{IS}$ for the coldest equilibrated state point. 
Open symbols are MD data calculated during a constant heating procedure 
starting from the IS configuration marked with $*$, as explained in the text.
Lines are PEL-EOS for equilibrium 
(full lines, Eq.~(\protect\ref{eq:componenti})) 
and for the heating procedure
(dashed lines, Eq.~(\protect\ref{eq:heating})).
\label{fig:pmdpel} 
}
\end{figure}
\begin{figure} 
\centering
\includegraphics[width=.45\textwidth]{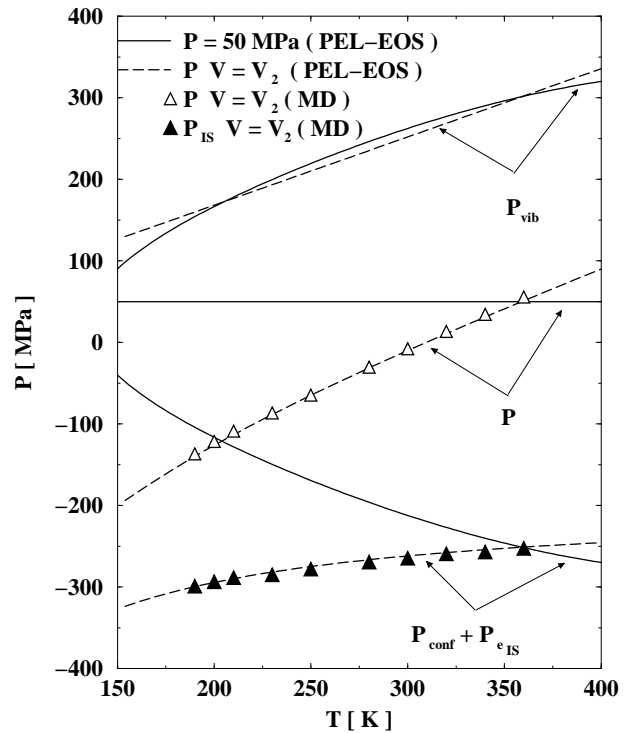}
\vspace{-0.5cm}
\caption{
Total ($P$), vibrational ($P_{vib}$) 
and $e_{IS}$ plus configurational ($P_{e_{IS}}+P_{conf}$) 
contributions to $P$ along constant $V$ (dashed lines)
and constant $P$ (solid lines) paths. At
constant $V$, $P_{vib}$ is linear in $T$ 
(Eq.~(\protect\ref{eq:pvib})).
\label{fig:constanV}
}
\end{figure}

The present approach also predicts the behavior
of $P$ when an IS configuration is heated from
$T=0$ at constant $V$.
While the system remains in the same basin, $P$ is
given by $P_{IS}(T_{IS},V) + P_{vib}(T,T_{IS},V)$ where
the only (linear) $T$-dependent part arises from

\begin{equation}
P_{vib}(T,T_{IS},V)= - k_B T \frac{\partial}{\partial V}  (a+b\,
e_{IS}(T_{IS})).
\label{eq:heating}
\end{equation}

We now apply the present derivation to 
the simple Lewis and Wanstr\"{o}m (LW) model for
the fragile molecular liquid 
orthoterphenyl (oTP)~\cite{mossa02,lewis}.
The LW model is a rigid three-site model, with intermolecular
site-site interactions described by the 
Lennard Jones (LJ) potential~\cite{lewis}.
Its simplicity allows one to reach simulation
times of the order of $\mu s$~\cite{mossa02}.
In this model, as in the LJ case, the anharmonic contributions are negligible,  
$e_{IS}(T)$ goes as $1/T$, and $\sum_{i=1}^{M} \ln(
\omega_i(e_{IS})/\omega_o)$ 
is linear in $e_{IS}$~\cite{mossa02}.  
Hence this model is a perfect candidate
for testing the validity of the PEL-EOS proposed here.

We use the excellent data base of state points
presented in Ref.~\cite{mossa02}  {\em i)} to calculate the $V$ dependence of 
the PEL parameters; {\em ii)} to derive the EOS for the oTP model, 
and {\em iii)} to compare it with the ``exact'' EOS calculated 
using the virial expression,  
as commonly implemented in molecular dynamics (MD) codes.

Fig.~\ref{fig:fit} shows, for five constant $V$ paths,
the simultaneous fit of the $T$ dependence of 
$S_{conf}$, the basin curvatures, and of $\langle
e_{IS}(T)\rangle$, according to Eqs.~(\ref{eq:sconf2}),
(\ref{eq:logw}), and (\ref{eq:eis}). 
The possibility of fitting simultaneously, with the same values of
$\alpha$, $\sigma$, $E_0$, $a$ and $b$, the quantities
$\langle e_{IS}(T)\rangle$, $\sum_{i=1}^{M}\ln( \omega_i(e_{IS})/\omega_o)$ 
and $S_{conf}(T)$, supports the validity of the two main assumptions, i.e.
Eq.~(\ref{eq:gdos}) and Eq.~(\ref{eq:fbasin2}).

The $V$ dependence of the three combinations of fit parameters,
$-e_{\infty}$, $S_{\infty}-k_B(a+be_{\infty})$ and 
$\sigma^2/2k_B$ is shown in  Fig.\ref{fig:fitp}. 
$P(V,T)$ can be calculated
from the $V$ derivative of these quantities 
according to Eq.~(\ref{eq:componenti}) and compared with the MD data. 
Such comparison is reported in  Fig.\ref{fig:pmdpel}. The 
very good agreement between the two set of data at all $V$'s 
confirms that a very accurate EOS based on PES properties 
has been derived for this oTP model.

The $T$ dependence of the individual
contributions can be evaluated according 
to Eqs.~(\ref{eq:pconf})-(\ref{eq:pvib}).
Fig.~\ref{fig:constanV} shows --- along a constant
$V$ and a constant $P$ path --- $P_{vib}$ and $P_{conf}+P_{e_{IS}}$.
We note that at constant $V$,
both $P$ components are increasing with $T$, while at constant $P$,
$P_{conf}+P_{e_{IS}}$ decreases on heating to compensate for the
increase of $P_{vib}$. 
 
The PES-EOS allows us also to contrast the isobaric and isochoric 
$T$ dependence of $S_{conf}$, $\langle e_{IS}(T)\rangle$ and 
$\sum_{i=1}^{M} \ln( \omega_i/\omega_o)$. Such comparison is
reported in Fig.~\ref{fig:fit}. Here we note the faster
decrease of $S_{conf}(T)$ and $\langle e_{IS}(T)\rangle$ 
along constant $P$ paths as well as the
different trend in the change of basin shape\cite{nota}. 

\begin{figure}[t] 
\centering
\includegraphics[width=.45\textwidth]{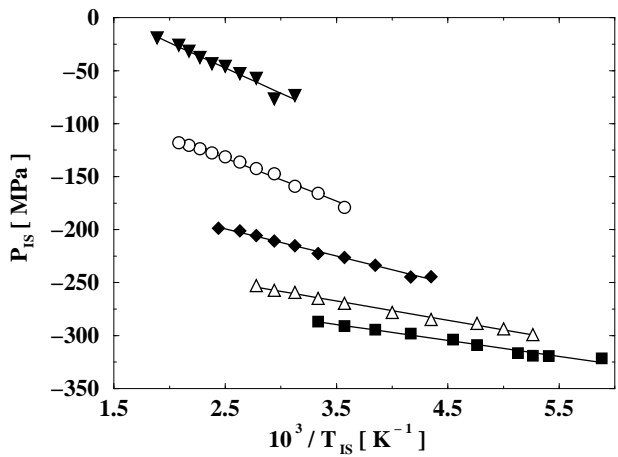}
\vspace{-0.5cm}
\caption{
Inherent structure equation of state. 
Symbols are MD calculations, lines are PEL predictions.
\label{fig:iseos}
}
\end{figure}

As a further check of the quality of the calculated EOS for the
oTP model, Fig.~\ref{fig:iseos} compares 
the MD (virial) and PEL (Eq.~(\ref{eq:pis})) IS-EOS. 
Since for the oTP model the term linear in $T$ in
Eq.~(\ref{eq:pis}) is negligible~\cite{lanave02}, 
$P_{IS}$ follows, to a good extent, a $1/T_{IS}$ law.
The quality of the comparison supports the interpretation of
$P_{IS}$ as the $V$ derivative of the depth and number of
basins sampled in the corresponding thermodynamic equilibrium state.

Finally, Fig.~\ref{fig:pmdpel} compares MD data (open symbols) 
and PEL (dashed lines) predictions in a run where 
$T$ is increased starting from the $T=0$ IS configuration,
as previously discussed. The entire simulation is much shorter than
the time needed to change basin, so that only the vibrational
degrees of freedom are thermalized. Also in this case, the PEL expression 
accounts for the observed linear increase of $P$. 

The present EOS,  based exclusively on PEL properties, can and will be used 
to address important issues in the thermodynamics of 
supercooled liquids~\cite{footnotebivariata}, such as the study of the
intrinsic limit of stability of the 
liquid state~\cite{debenedettieos,roberts,sastry00} and the IS-based 
thermodynamic description of aging~\cite{aging}.
%
%
%

%
%

\begin{thebibliography}{99}
\bibitem{anomali} 
Note that liquids with non monotonic behavior of density
(as water) must be characterized by  a positive $P_{1/T}$ coefficient.
\bibitem{footnotebivariata} 
Another interesting possibility
is offered by a generalization to $V$ of Eq.~(\protect\ref{eq:gdos}).
The oTP data allow us to eliminate the simple possibility that
$\Omega(e_{IS},V)$ is a bivariate Gaussian~\protect\cite{lanave02}.
\end{thebibliography}
\end{document}